\documentclass[a4paper]{article}

\usepackage{philex}
\usepackage{amssymb}
\usepackage{latexsym}
\usepackage{bussproofs}
\usepackage{pdflscape}
\usepackage{url}

\usepackage{natbib}

\usepackage{mathabx}
\usepackage{stackengine}
\stackMath

\newtheorem{definition}{Definition}

\newtheorem{theorem}{Theorem}

\usepackage{pxfonts}
\usepackage{microtype}

\title{Definite Descriptions in Intuitionist Positive Free Logic}
\author{Nils K\"urbis}
\date{}

\begin{document}
\maketitle

\begin{center}
Published in \emph{Logic and Logical Philosophy}  30/2 (2021): 327-358\\
\url{http://dx.doi.org/10.12775/LLP.2020.024}\bigskip
\end{center}

\begin{abstract}
\noindent This paper presents rules of inference for a binary quantifier $I$ for the formalisation of sentences containing definite descriptions within intuitionist positive free logic. $I$ binds one variable and forms a formula from two formulas. $Ix[F, G]$ means `The $F$ is $G$'. The system is shown to have desirable proof-theoretic properties: it is proved that deductions in it can be brought into normal form. The discussion is rounded up by comparisons between the approach to the formalisation of definite descriptions recommended here and the more usual approach that uses a term-forming operator $\iota$, where $\iota xF$ means `the F'. 
\end{abstract}

\noindent Keywords: free logic, definite descriptions, proof theory, normalisation, intuitionist logic, binary quantifiers, term forming operators

\section{Introduction} 
In two recent papers, I presented a binary quantifier for the formalisation of definite descriptions, which was added to a system of natural deduction for an intuitionist negative free logic.\footnote{See \citep{kurbisiotaI} and \citep{kurbisiotaII}. I shall divert from the previous presentation in two respects. First, whereas previously I used $\iota$ for the binary quantifier, I now use $I$ to make the distinction between it and a term-forming operator for definite descriptions more perspicuous. Secondly, the formalisation of quantificational logic uses parameters for free variables.} In the simplest case, $I$ forms a formula from two predicates, binding a variable. For example, if $F$ is `$x$ is present King of France' and $G$ is `$x$ is bald', then $Ix[F, G]$ means `The present King of France is bald'. In the general case, formulas of any complexity may take the place of $F$ and $G$.\footnote{To avoid vacuous quantification, we could require these to be formulas that contain the variable $x$ free. But this is not necessary.} I shall use $F$ and $G$ for any formulas where the formalisation of definite descriptions is concerned. $A$ and $B$ are used for formulas in general. Where this aids the discussion, the occurrence of a free variable in a formula is indicated by enclosing it in brackets following the formula, as in $F(x)$ and $G(x)$, and replacement of variables by terms will be indicated analogously, as in $F(t)$ and $G(t)$. In official notation, $A^x_t$ denotes the result of replacing the variable $x$ by the term $t$ in the formula $A$. The syntax of $I$ is that if $F$ and $G$ are formulas and $x$ is a variable, then $Ix[F, G]$ is a formula in which $x$ is bound. Its intended meaning is `The $F$ is $G$'. 

In negative free logic, the meaning of `The $F$ is $G$' is given by its Russellian analysis `There is exactly one $F$ and it is $G$', and accordingly, $Ix[F, G]$ is equivalent to $\exists x (F\land\forall y(F_y^x\rightarrow y=x)\land G)$. Definite descriptions are therefore eliminable. For this reason many free logicians prefer positive over negative free logic. In positive free logic, the Russellian analysis of `The $F$ is $G$' is rejected and only one half of the equivalence holds: if there is exactly one $F$ which is $G$, then the $F$ is $G$, but not conversely. Positive and negative free logicians agree, however, that `The $F$ exists' is equivalent to `There is exactly one $F$', and both have an equivalent formalisation in $\exists y\forall x(F\leftrightarrow x=y)$. 

The present paper investigates which rules for the binary quantifier $I$ are suitable additions to a positive intuitionist free logic. In preparation for this task the more common approach to formalising definite descriptions within free logic is presented, which uses a term-forming operator $\iota$: $\iota$ binds a variable and forms a singular term out of a formula, where $\iota xF$ means `the $F$'. `The $F$ is $G$' is formalised as $G(\iota xFx)$. Establishing some of the logical properties of formulas of the form $G(\iota xF)$ within intuitionist positive free logic presents a vital step towards the formulation of rules for the binary quantifier $I$. Let me say already here that, due to the characteristics of positive free logic, these rules are significantly more complex than those for $I$ in negative free logic. 

The rules for $I$ presented in the previous papers have desirable proof-theoretic properties: a normalisation theorem showed that formula occurrences that are the conclusions of an introduction rule and major premises of an elimination rule for their main connective can be removed from deductions in the system of intuitionist negative free logic with the binary quantifier $I$. Following Dummett and Prawitz, the rules for $I$ are \emph{in harmony} and thus can count as specifying its meaning. The present paper follows a similar path. The investigation is proof-theoretical and a normalisation theorem is established for the system of positive intuitionist free logic extended by rules for the binary quantifier $I$.\footnote{For the philosophical importance of the normalisation of deductions, see \citep{dummettjustdeduction}, \citep[Chs. 10-13]{dummettLBM}, \citep{prawitzdummett}, \citep{prawitzmeaningviaproofs}. For a brief overview of the motivations behind, challenges to and prospects for Dummett's and Prawitz's approach, see \citep{kurbisPTSNM}. Normalisation for classical and intuitionist logic was first proved by Prawitz \citep{prawitznaturaldeduction}.}

Despite its proof-theoretic stance, occasionally the present paper touches upon semantical considerations to illustrate and motivate the use of the binary quantifier $I$. These remain at an intuitive level. The semantic intuition behind adopting a positive free logic is that atomic sentences containing terms that do not refer (to an object considered to exist or to be in the domain of quantification) may nonetheless be true. This opposes the Russellian analysis according to which they are all false. The failure of the Russellian equivalence of `The $F$ is $G$' and `There is exactly one $F$ and it is $G$' allows for the possibility that the $F$ is $G$ even though the $F$ does not exist. Thus according to positive free logic, it may be true that John admires the world's most famous detective even though the world's most famous detective, Sherlock Holmes, does not exist. Furthermore, the law of self-identity holds unrestrictedly and not just for terms that refer, so that `Sherlock Holmes is identical to Sherlock Holmes' is logically true. 

Formalising sentences containing definite descriptions with the binary quantifier $I$ has certain advantages over the more usual approach that employs the forming operator $\iota$. In the latter, $\iota xF$ denotes the only $F$, if there is one, or else an object not considered to be amongst those that exist or nothing at all. A question arises concerning the scope of unary operators, here only negation, but in a modal setting also the modal operators: what does $\neg G(\iota x F)$ mean? Does it mean that the $F$ is not $G$ or that it is not the case that the $F$ is $G$? 

There is a sense in which no decision is called for. In negative free logic, $G(\iota xF)$ is true just in case there is a unique $F$ and it is $G$. In positive free logic $G(\iota xF)$ is true just in case there is a unique $F$ and it is $G$ or the object assigned to $\iota xF$ that is not considered to be amongst those that exist is $G$. Thus assuming the principle of bivalence, as many prominent free logicians do, in negative free logic, $\neg G(\iota x F)$ is true just in case either there is no unique $F$ or there is a unique $F$ and it is not $G$; in positive free logic, $\neg G(\iota x F)$ is true if either there is a unique $F$ and it is not $G$ or if there is no unique $F$ and the `non-existent' object assigned to $\iota xF$ is not $G$. In the negative setting, there is a formula equivalent to $\neg G(\iota xF)$ in the language that does not contain $\iota$ and conveniently displays its truth conditions; in the positive setting, there is no such formula, as there is no other way of expressing that the `non-existent' object assigned to $\iota xF$ is not $G$ than $\neg G(\iota xF)$. In either case, $\neg G(\iota xF)$ has disjunctive truth conditions. 

Although there is nothing wrong with disjunctive truth conditions -- disjunctions, after all, have them and are on the whole well understood -- as the discussion shows we can evidently draw a distinction between the \emph{internal} negation `The $F$ is not $G$' and the \emph{external} negation `It is not the case that the $F$ is $G$' of `The $F$ is $G$' and a need to do so often arises. Hence it is desirable to have the means to express the distinction in the formal language. For that purpose, it is necessary to introduce markers for \emph{scope distinctions}. Many authors, especially those working with definite descriptions in modal logic, introduce an operator $\lambda$ for predicate abstraction for that purpose.\footnote{See \citep{lambertfreedef} for a treatment of definite descriptions and scope distinctions with an abstraction operator in classical negative free logic, and \citep[Chs. 9ff]{mendelsohnfitting} and \citep[Ch. 19]{garsonmodallogic} for the same in modal extensions of classical positive free logic. \cite{lambertbencivengacomplexpred} formalise a system of classical positive free logic with the $\lambda$ abstraction operator, but without definite descriptions. A different approach is followed by Gratzl, who employs Russell's method of marking the scope of a definite description by repeating it in square brackets \citep{gratzldefdescr}.} The binary quantifier $I$, by contrast, has scope distinctions built directly into the notation: the internal negation `The $F$ is not $G$' is formalised as $Ix[F, \neg G]$, the external negation `It is not the case that the $F$ is $G$' as $\neg Ix[F, G]$. There is thus no need for separate syntactic means to mark scope distinctions and using the binary quantifier to formalise definite descriptions allows for a certain economy in the language.\footnote{This is not to say that the binary quantifier only has advantages over the term-forming operator. Although recommended by Dummett \citep[p.162]{dummettfregelanguage}, it is not recommended by Bostock \citep[Sec. 8.4]{bostockintermediate}. It is fair to say that formulas with nested binary quantifiers can be difficult to read, but such complications are unavoidable where scope distinctions are to be considered.}

Section 6 of this paper contains a formal comparison of the present system with a system using a term-forming operator $\iota$ for definite descriptions. But first, the next section presents the system of intuitionist positive free logic used in the present paper.\footnote{It is the positive version of the system of intuitionist negative free logic of \citep{kurbisiotaI}, and thus an intuitionist version of a standard system of classical positive free logic used by Lambert, Bencivenga and others. Indrzejczak has shown how to formalise a variety of classical and intuitionist negative and positive free logic in sequent calculi that allow for cut elimination \citep{andrzejcutfreefreelogic}.}

\section{Intuitionist Positive Free Logic} 
The language of the system \textbf{IPF} of intuitionist positive free logic is standard. For simplicity I assume that there are no function symbols. The terms of the language are constant symbols and parameters for free variables. $t$ is used for terms of either kind. Whenever the term-forming operator $\iota$ is concerned in later sections, $t$ also ranges over definite descriptions, unless otherwise stated.

Square brackets around top-most formulas in deductions indicate assumption classes. Every formula occurrence in a deduction that is not regarded as an axiom is in some assumption class, where formulas of different type are in different classes, and those of the same type may or may not be in the same class. Every assumption class receives a label. Discharge or closing of assumptions is indicated by repeating the label at the inferences where the formula occurrences in the assumption class are discharged. Empty assumption classes are allowed and used in vacuous discharge. For convenience I will only display the assumption classes of discharged assumptions in deductions given below, and adopt the convention that undischarged assumptions of the same type belong to the same assumption class. 

The rules for the propositional connectives are: 

\begin{center}
\bottomAlignProof
\AxiomC{$A$} 
\AxiomC{$B$}
\LeftLabel{$\land  I$: \ }
\BinaryInfC{$A\land  B$}
\DisplayProof\qquad\qquad 
\bottomAlignProof
\AxiomC{$A\land  B$}
\LeftLabel{$\land  E$: \ }
\UnaryInfC{$A$}
\DisplayProof\qquad 
\bottomAlignProof
\AxiomC{$A\land  B$} 
\UnaryInfC{$B$}
\DisplayProof
\end{center}

\begin{center}
\bottomAlignProof
\AxiomC{$[A]^i$}
\noLine
\UnaryInfC{$\Pi$}
\noLine
\UnaryInfC{$B$}
\RightLabel{$_i$}
\LeftLabel{$\rightarrow I$: \ }
\UnaryInfC{$A\rightarrow B$}
\DisplayProof \qquad \qquad
\bottomAlignProof
\AxiomC{$A\rightarrow B$}
\AxiomC{$A$}
\LeftLabel{$\rightarrow E$: \ }
\BinaryInfC{$B$}
\DisplayProof
\end{center}

\begin{center}
\bottomAlignProof
\AxiomC{$A$}
\LeftLabel{$\lor I$: \ }
\UnaryInfC{$A\lor B$}
\DisplayProof\qquad
\bottomAlignProof
\AxiomC{$B$}
\UnaryInfC{$A\lor B$}
\DisplayProof\qquad\qquad
\bottomAlignProof
\AxiomC{$A\lor B$}
\AxiomC{$[A]^i$}
\noLine
\UnaryInfC{$\Pi$}
\noLine
\UnaryInfC{$C$}
\AxiomC{$[B]^j$}
\noLine
\UnaryInfC{$\Sigma$}
\noLine
\UnaryInfC{$C$}
\RightLabel{$_{i, j}$}
\LeftLabel{$\lor E$: \ }
\TrinaryInfC{$C$}
\DisplayProof
\end{center}

\begin{prooftree}
\AxiomC{$\bot$}
\LeftLabel{$\bot E$: \ }
\UnaryInfC{$B$}
\end{prooftree}

\noindent where the conclusion $B$ of $\bot E$ is restricted to atomic formulas. 

The rules for the quantifiers appeal to a primitive predicate $\exists !$, to be interpreted as `exists' or `refers':\bigskip 

\begin{center}
\bottomAlignProof
\AxiomC{$[\exists !a]^i$}
\noLine
\UnaryInfC{$\Pi$}
\noLine
\UnaryInfC{$A_a^x$}
\LeftLabel{$\forall I:$ \ }
\RightLabel{$_i$}
\UnaryInfC{$\forall x A$}
\DisplayProof\qquad\qquad
\bottomAlignProof
\AxiomC{$\forall xA$}
\AxiomC{$\exists !t$}
\LeftLabel{$\forall E:$ \ }
\BinaryInfC{$A_t^x$}
\DisplayProof
\end{center} 

\noindent where in $\forall I$, $a$ does not occur in $A$ nor in any undischarged assumptions of $\Pi$ except $\exists !a$. 

\begin{center}
\bottomAlignProof
\AxiomC{$A_t^x$}
\AxiomC{$\exists !t$}
\LeftLabel{$\exists I:$ \ }
\BinaryInfC{$\exists x A$}
\DisplayProof\qquad\qquad
\bottomAlignProof
\AxiomC{$\exists xA$}
\AxiomC{$\underbrace{[A_a^x]^i, \ [\exists !a]^j}$}
\noLine
\UnaryInfC{$\Pi$}
\noLine
\UnaryInfC{$C$}
\RightLabel{$_{i, j}$}
\LeftLabel{$\exists E:$ \ }
\BinaryInfC{$C$}
\DisplayProof
\end{center} 

\noindent where in $\exists E$, $a$ does not occur in $A$ nor in $C$ nor in any undischarged assumptions of $\Pi$, except $A_a^x$ and $\exists ! a$. 

Identity is governed by the Law of Self-Identity and Leibniz' Law:  

\begin{center}
$=I$: \ $t=t$
\qquad\qquad
\AxiomC{$t_1=t_2$}
\AxiomC{$A_{t_1}^x$}
\LeftLabel{$= E$: \ } 
\BinaryInfC{$A_{t_2}^x$}
\DisplayProof
\end{center} 

\noindent where $A$ is an atomic formula. The general case is proved by induction over the complexity of formulas. Requiring $t_1$ and $t_2$ to be different excludes vacuous applications of $=E$. This is obviously no restriction and absolves us from considering maximal formulas of the form $t_1=t_2$. In the unfortunate circumstances where a transformation of a deduction results in a vacuous application of $=E$, it is assumed that it is removed as part of the transformation. 

The characteristic difference between positive and negative free logic is that the former lacks the \emph{rules of strictness}, which allow the derivation of $\exists !t$ from atomic formulas containing $t$, and the latter requires the premise $\exists !t$ in the introduction rule for identity $= I^n$. Intuitively, this means that in positive free logic, atomic sentences may be true even if terms occurring in them do not refer, while such sentences are always false in negative free logic; and in negative free logic, self-identity is equivalent to existence, while in positive free logic, this is not in general the case. 

It will make sense to have primitive rules for the biconditional $\leftrightarrow$:\bigskip 

\begin{center}
\AxiomC{$[A]^i$}
\noLine
\UnaryInfC{$\Pi$}
\noLine
\UnaryInfC{$B$}
\AxiomC{$[B]^j$}
\noLine
\UnaryInfC{$\Pi$}
\noLine
\UnaryInfC{$A$}
\LeftLabel{$\leftrightarrow I: \ $}
\RightLabel{$_{i, j}$}
\BinaryInfC{$A\leftrightarrow B$}
\DisplayProof

\bigskip 

\AxiomC{$A\leftrightarrow B$}
\AxiomC{$A$}
\LeftLabel{$\leftrightarrow E^1: \ $}
\BinaryInfC{$B$}
\DisplayProof\quad
\AxiomC{$A\leftrightarrow B$}
\AxiomC{$B$}
\LeftLabel{$\leftrightarrow E^2: \ $}
\BinaryInfC{$A$}
\DisplayProof
\end{center} 

\noindent Having these rules available simplifies some of the deductions in the comparison of the binary quantifier and term-forming operator for definite descriptions. 

It is customary in discussions of classical free logic to observe that it is possible to dispense with the primitive $\exists !$ and to treat it as defined in terms of $\exists$ and $=$, due to the following equivalence which holds also in $\mathbf{IPF}$: 

\lbp{1}{$\ast 1$}{$\vdash \exists !t\leftrightarrow \exists y \ y=t$} 

\noindent If $\exists ! t$, then by $=I$ and $\exists I$: $\exists y \ y=t$. If $\exists y \ y=t$, suppose $a=t$ and $\exists ! a$, then by $=E$: $\exists ! t$, and so by $\exists E$: $\exists ! t$.\bigskip 

\noindent It is, however, formally convenient and philosophically preferable to keep $\exists !$ primitive. From the formal perspective of proving a normalisation theorem for $\mathbf{IPF}$, if $\exists !$ is treated as defined, we should have to consider additional maximal formulas of the form $\exists y \ y=t$ arising when the premise $\exists !t$, i.e. $\exists y \ y=t$, of $\exists I$ or $\forall E$ is derived by $\exists I$. From the philosophical perspective, and one that aims at a proof-theoretic specification of the meanings of the logical constants, if $\exists !$ is treated as defined, then the meaning of the universal quantifier would be specified in terms of the existential quantifier, that expression appearing in a premise of its introduction rule and discharged hypothesis of its elimination rule, and the specification of the meaning of the existential quantifier would be circular for precisely the same reason.  

Wider philosophical considerations also support the view that $\exists !$ should be regarded as primitive. Logic is silent about the meanings of the atomic sentences of the formal language and is not concerned with the question which of them are, in fact, true and which are false. To treat $\exists!$ as primitive is to treat formulas of the form $\exists !t$ as being in the same category as the other atomic formulas of the language in this respect: the intended interpretation of $\exists!$ is given outside logic. Notice that the formal system does not decide whether $\exists !t$ should mean `$t$ exists' or `$"t"$ refers'. What exists and what doesn't, which terms refer and which don't, is not in general a question of logic---although it may be in very special cases, such as the numbers for logicists. Logic has no say in what grounds the rather crucial difference between the names `Jamina' and `Pegasus'. The first refers, the second does not. Jamina is a pygmy hippopotamus that lives in \L\'od\'z Zoo. Pegasus is a winged horse from Greek mythology. Jamina exists, Pegasus does not. `$\exists !$Jamina' is true in virtue of the animals that are kept in \L\'od\'z Zoo and the names they have been given, and not because of some feature of logic. The existence of Pegasus or the assumption that `Pegasus' refers stands in contradiction with what there is, and maybe even with what there can be according to the laws of biology, and hence $\neg\exists !$Pegasus, but what it is that precludes that Pegasus exists or that `Pegasus' refers is again not a question of logic. 

These philosophical considerations also accord with the fact that the use of $\exists !$ in deductions in positive free logic is rather limited. No rule of $\mathbf{IPF}$ has a formula of the form $\exists ! t$ as the conclusion. There is no introduction rule for $\exists !$. Its main use is as an assumption, possibly one to be discharged. As a corollary of the normalisation theorem for $\mathbf{IPF}$ it could be established rigorously under which conditions a formula of the form $\exists !t$ may be derived. This issue is tangential to the main concerns of this paper, and so to do so here would go too far. For the present discussion it suffices to point out that, aside from by being an assumption, a subformula of one or a consequence of $\bot$, a formula of the form $\exists !t$ may be derived by Leibniz' Law, in which case it has been derived from a formula of the same form, but containing a different term, and an identity with $t$ to its left. Assumptions of the latter kind may be discharged by an application of $\exists E$, as exemplified by the proof of the right to left direction of \rf{1}, which presents the only other way a formula of the form $\exists !t$ may be derived from premises that are consistent and do not contain it as a subformula. Thus formulas of the form $\exists !t$ are only derivable if the premises are inconsistent, the formula occurs as a subformula amongst them, or they contain $\exists$ or identities, the term $t$ and the existential quantifier. Accordingly, in applications of the system of positive free logic in the formalisation of a theory, for instance, or of ordinary argumentation, typically a stock of formulas would be given that specify what exists or which terms refer. The logic alone does not allow us to derive a formula of the form $\exists ! t$, but this requires assumptions which are of a non-logical character. In the case of negative free logic, there are also the rules of strictness that allow the derivation of formulas of the form $\exists !t$, but to do so, assumptions or axioms need to be given that are atomic formulas, and their truth is again not established by logic.   

Consequently, a derivation of `The $F$ is $G$' requires information that is not of a purely logical kind, be it that these are assumptions of a deduction or non-logical axioms of a theory. For instance, as will be seen from the introduction rule governing $I$ to be given in section 4, if we are to derive that the female pygmy hippopotamus of \L\'od\'z Zoo is hungry, then we require assumptions such as that Jamina is a female pygmy hippopotamus of \L\'od\'z Zoo, that she exists, that she is hungry, and that she is the only female pygmy hippopotamus of \L\'od\'z Zoo. Positive free logic also allows for the option that a sentence `The $F$ is $G$' is true even if there is no unique $F$, but then the logic does not specify any conditions under which this may be the case: rather, this depends entirely on the non-logical content of $F$ and $G$. Such sentences can only be used as assumptions or need to be added as non-logical axioms to theories. In this respect, however, they are no different from axioms of theories such as that every number has a unique successor (\emph{pace} logicists) or assumptions such as that there is only one female pygmy hippopotamus in \L\'od\'z Zoo. 

The discussion of the previous paragraphs hints at a stronger conclusion: the meaning of $\exists!$ cannot be given by rules of inference, at least not by rules of a purely logical character. This is neither surprising nor problematic. Existence and reference concern domains outside logic, and there are other ways of giving meanings to expressions than laying down rules of inference for them. Logic can rely on those for the meaning of $\exists!$, just as it relies on them to provide the meanings of `is a pygmy hippopotamus' and `is a winged horse'. This may be contentious amongst inferentialists, and as such point to a problem for the acceptability of free logics with an existence predicate to some of them, but to address these issues would require a paper on its own.\footnote{In negative free logic, $\exists!$ is governed by what may look like introduction rules, namely the rules of strictness. Even should we treat them as such, $\exists!$ has no elimination rules, and so the rules governing it fail to exhibit the format required of rules that determine meanings. Should we treat the rules for the quantifiers and $=I^n$ as the elimination rules for $\exists!$, then the rules governing it are not harmonious, and so again its meaning is not determined by them.}

\section{The Term-Forming $\iota$ Operator in IPF}
The binary quantifier $I$ is intended to formalise the ordinary English `The $F$ is $G$'. In formalisations of definite descriptions by a term-forming operator $\iota$, it is customary to provide axioms only for occurrences of $\iota$-terms to the left or right of identity and to let the logical properties of formulas with occurrences of $\iota$-terms in other contexts be determined by them. As a preparation for finding suitable rules of natural deduction governing the binary quantifier $I$, in this section I will consider adding a term-forming operator $\iota$ for definite descriptions to \textbf{IPF} and investigate the properties of the more general formulas of the form $G(\iota xFx)$, where $G$ need not be identity, in the resulting system. 

The Russellian analysis of `The $F$ is $G$' as `There is a unique $F$ and it is $G$' is not suitable for the framework of positive free logic, where, semantically speaking, atomic formulas may be true even though they contain terms that do not refer (to an object considered to exist or to be in the domain of quantification). $G(\iota xF)$ is not logically equivalent to $\exists y (\forall x(F\leftrightarrow x=y)\land G)$ for every choice of $G$. If there is a unique $F$ and it is $G$, then the $F$ is $G$, but the converse holds only under the condition that there is a unique $F$. If $\exists y\forall x(F\leftrightarrow x=y)\land G(\iota xF)$, then $\exists y (\forall x(F\leftrightarrow x=y)\land G)$, and if $\exists y (\forall x(F\leftrightarrow x=y)\land G)$, then $G(\iota xF)$. But $\exists y (\forall x(F\leftrightarrow x=y)\land G)$ also implies $\exists y\forall x(F\leftrightarrow x=y)$. So in positive free logic, $G(\iota xF)\land \exists y \forall x(F\leftrightarrow x=y)$ is equivalent to $\exists y (\forall x(F\leftrightarrow x=y)\land G)$. More briefly, exploiting the equivalence between `the $F$ exists' and `there is a unique $F$', which is retained in positive free logic, $G(\iota xF)\land \exists !\iota xF$ is equivalent to $\exists y (\forall x(F\leftrightarrow x=y)\land G)$. 

To establish the observations of the previous paragraph formally in \textbf{IPF} extended by the term-forming operator for definite descriptions, we add \emph{Lambert's Law} as the sole axiom governing $\iota$: 

\lbp{LL}{$LL$}{$\forall y(\iota xF=y \leftrightarrow \forall x(F\leftrightarrow x=y))$} 

\noindent where $x$ and $y$ are distinct.\bigskip

\noindent Call the resulting system $\mathbf{IPF}^\iota$. 

It is generally agreed that Lambert's Law axiomatises the minimal theory of $\iota$. Added to classical positive free logic, the resulting theory is often called $FD$ or $MFD$.\footnote{See \citep{fraassenlambert}, \citep{bencivengahandbook}, \citep{lambertfreedef}.} I am here interested in formalising an equally minimal theory of the binary quantifier $I$. As the chosen deductive apparatus of the present paper is natural deduction, and the aim is to formulate rules with satisfactory proof-theoretic properties, the logic is intuitionist. However, the rules for $I$ to be given in the next section could equally be added to classical positive free logic.\footnote{Rules adequate for the binary quantifier $I$ in a classical sequent calculus that allow cut elimination are the subject of another paper. Indrzejczak provided cut-free sequent calculi for various formalisations of the term-forming operator $\iota$ (\citep{andrzejmodaldescription}, \citep{andrzejfregean}, \citep{andrzejfreedefdescr}). Czermak formalised a further cut free system for a logic of definite descriptions \citep{czermakdefdescr}. Tennant provides normalising rules for $\iota$ that are equivalent to Lambert's Law in intuitionist negative free logic \citep{tennantnatural}, \citep{tennantabstraction}.}

It is easy to show that \rf{LL} implies that there is a unique $F$ if and only if the $F$ exists: 

\lbp{2}{$\ast 2$}{$\vdash\exists !\iota xF\leftrightarrow \exists y \forall x(F\leftrightarrow x=y)$}

\noindent (a) By \rf{LL} and $\forall E$: $\exists !\iota xF\vdash  \iota xF=\iota xF \leftrightarrow \forall x(F\leftrightarrow x=\iota xF)$, so by $=I$ and $\leftrightarrow E$: $\exists !\iota xF\vdash\forall x(F\leftrightarrow x=\iota xF))$, so by $\exists I$: $\exists !\iota xF\vdash \exists y \forall x(F\leftrightarrow x=y)$. 

\noindent (b) Conversely, assume $\exists y \forall x(F\leftrightarrow x=y)$, and suppose $\exists !a$ and $\forall x(F\leftrightarrow x=a)$. Then by \rf{LL} and $\forall E$: $\iota xF=a$. So by $\exists I$: $\exists y \ \iota xF=y$, and so by \rf{1}: $\exists !\iota xF$.\bigskip

\noindent Next, if the $F$ exists and it is $G$, then there is a unique $F$ that is $G$: 

\lbp{3}{$\ast 3$}{$\exists !\iota xF, G(\iota x F)\vdash \exists y(\forall x(F\leftrightarrow x=y)\land G(y))$}

\begin{prooftree}
\AxiomC{$\iota xF=\iota xF$}
\AxiomC{$(LL)$}
\AxiomC{$\exists ! \iota xF$}
\BinaryInfC{$\iota xF=\iota xF \leftrightarrow \forall x(F\leftrightarrow x=\iota xF)$}
\BinaryInfC{$\forall x(F\leftrightarrow x=\iota xF)$}
\AxiomC{$G(\iota xF)$}
\BinaryInfC{$\forall x(F\leftrightarrow x=\iota xF)\land G(\iota xF)$}
\AxiomC{$\exists !\iota xF$}
\BinaryInfC{$\exists y(\forall x(F\leftrightarrow x=y)\land G(y))$}
\end{prooftree} 

\begin{landscape}
\noindent \rf{LL} implies that if there is a unique $F$ that is $G$, then the $F$ is $G$:

\lbp{4}{$\ast 4$}{$\exists y(\forall x(F\leftrightarrow x=y)\land G(y))\vdash G(\iota x F)$}

\begin{prooftree}
\AxiomC{$\exists y(\forall x(F\leftrightarrow x=y)\land G(y))$}
\AxiomC{$(LL)$}
\AxiomC{$[\exists ! a]^1$}
\BinaryInfC{$\iota xF=a \leftrightarrow \forall x(F\leftrightarrow x=a)$}
\AxiomC{$[\forall x(F\leftrightarrow x=a)\land G(a)]^2$}
\UnaryInfC{$\forall x(F\leftrightarrow x=a)$}
\BinaryInfC{ $\iota xF=a$}
\AxiomC{$[\forall x(F\leftrightarrow x=a)\land G(a)]^2$}
\UnaryInfC{$G(a)$}
\BinaryInfC{$G(\iota xF)$}
\RightLabel{$_{1, 2}$}
\BinaryInfC{$G(\iota xF)$}
\end{prooftree} 

\noindent Conversely, \rf{LL} is derivable from \rf{3} and \rf{4}. Let \rf{3}' and \rf{4}' be \rf{3} and \rf{4} with $G(y)$ replaced by $\iota xF=y$: 

\lbp{5}{$\ast 5$}{\rf{3}', \rf{4}' $\vdash$ \rf{LL}}

\begin{prooftree} 
\AxiomC{$[\iota xF=a]^2$}
\AxiomC{$[\iota xF=a]^2$}
\AxiomC{$[\exists !a]^4$}
\BinaryInfC{$\exists ! \iota xF$}
\RightLabel{$_{(\ast 3)'}$}
\BinaryInfC{$\exists y(\forall x(F\leftrightarrow x=y)\land y=a)$}
\AxiomC{$[\forall x(F\leftrightarrow x=b)\land b=a]^1$}
\doubleLine
\UnaryInfC{$\forall x(F\leftrightarrow x=a)$} 
\RightLabel{$_1$}
\BinaryInfC{$\forall x(F\leftrightarrow x=a)$}
\AxiomC{$[\forall x(F\leftrightarrow x=a)]^4$} 
\AxiomC{$a=a$}
\BinaryInfC{$\forall x(F\leftrightarrow x=a)\land a=a$}
\AxiomC{$[\exists ! a]^4$}
\BinaryInfC{$\exists y(\forall x(F\leftrightarrow x=y)\land y=a)$}
\RightLabel{$_{(\ast 4)'}$}
\UnaryInfC{$\iota xF=a$}
\RightLabel{$_{2, 3}$}
\BinaryInfC{$\iota xF=a\leftrightarrow \forall x(F\leftrightarrow x=a)$} 
\RightLabel{$_4$}
\UnaryInfC{$\forall y( \iota xF=y\leftrightarrow \forall x(F\leftrightarrow x=y))$} 
\end{prooftree} 

\noindent The double line stands for the derivable inference $\forall x(F\leftrightarrow x=b)\land b=a\vdash \forall x(F\leftrightarrow x=a)$, and the labels of \rf{3}' and \rf{4}' indicate their use in the deduction.\bigskip

\noindent It follows that the minimal theory of definite descriptions can be axiomatised equivalently by \rf{3} and \rf{4} instead of \rf{LL}. To formalise suitable rules for the binary quantifier $I$, in the next section we will cast \rf{3} and \rf{4} into rules of a system of natural deduction. 
\end{landscape}

Russell and Whitehead observe that ambiguity arises from an attempted definition of $G(\iota xF)$ by $\exists y(\forall x(F\leftrightarrow x=y)\land G(y))$ `when $\iota xF$ occurs in a proposition which is part of a larger proposition': then `there is doubt whether the smaller or the larger proposition is to be taken as the $G(\iota xF)$.' (\citep[173]{PMto56}, notation adjusted.) They note that the formula  $G(\iota xF)\rightarrow B$, $B$ not containing the variable that $\iota xF$ replaces in $G$, can mean either of these:\bigskip

$\exists y(\forall x(F\leftrightarrow x=y)\land G(y))\rightarrow B$

$\exists y(\forall x(F\leftrightarrow x=y)\land (G(y)\rightarrow B))$\bigskip

\noindent Replacing $B$ with $\bot$, the distinction is the one between \emph{internal} and \emph{external} negation. If there is no unique $F$, then in negative free logic the first is true, but the second is false. There is, therefore, a need to distinguish them. Russell and Whitehead repeat the description in square brackets to mark scope distinctions:\bigskip

$[\iota xF]G(\iota xF)\rightarrow B$ 

$[\iota xF](G(\iota xF)\rightarrow B)$\bigskip

\noindent In the first, the description has narrow scope, in the second it has wide scope. 

In formalisations of theories of definite descriptions axiomatised by \rf{LL}, $G(\iota xF)$ is not defined as $\exists y(\forall x(F\leftrightarrow x=y)\land G(y))$. The logical properties of $G(\iota xF)$ are treated entirely in terms of consequences of \rf{LL}. As the latter makes no provision for scope distinctions, neither does the theory as a whole. 

These theories can be axiomatised equivalently by \rf{3} and \rf{4}. \rf{4} specifies the conditions under which $G(\iota xF)$ may be inferred and \rf{3} the consequences that follow from it. Adopting an inferentialist theory of meaning, these determine the meaning of $G(\iota xF)$. One may now ask what they tell us about $\neg G(\iota xF)$. There are two paths to addressing this question. One is to contrapose \rf{3} and \rf{4}, the other to replace to replace $G$ by $\neg G$ in them. Doing so with \rf{3}, keeping $\exists !\iota xF$ in the antecedent in the contraposition, yields:  

\lbp{8}{$\ast 6$}{$\neg \exists y(\forall x(F\leftrightarrow x=y)\land G(y)), \exists !\iota xF\vdash\neg G(\iota x F)$}
 
\lbp{9}{$\ast 7$}{$\neg G(\iota x F), \exists !\iota xF\vdash \exists y(\forall x(F\leftrightarrow x=y)\land \neg G(y))$}

\noindent \rf{8} and \rf{9} give grounds and consequences of $\neg G(\iota xF)$ under the assumption that the $F$ exists, in which case Russell's internal and external negations are equivalent. Contraposing and replacing in \rf{4} give grounds and consequences of $\neg G(\iota xF)$ independently of this assumption: 

\lbp{6}{$\ast 8$}{$\exists y(\forall x(F\leftrightarrow x=y)\land \neg G(y)\vdash \neg G(\iota x F)$}

\lbp{7}{$\ast 9$}{$\neg G(\iota x F)\vdash \neg \exists y(\forall x(F\leftrightarrow x=y)\land G(y)$}

\noindent $\neg G(\iota xF)$ thus lies between internal and external negation in logical strength, implied by the former and implying the latter, but equivalent to neither. Thus in a theory of definite descriptions axiomatised by \rf{LL}, $\neg G(\iota xF)$ presents a third option besides the two countenanced by Russell and Whitehead, and one might even say that such a theory endorses the ambiguity they diagnose in $\neg G(\iota xF)$. 

Russell and Whitehead have a point when they insist that it is possible to draw scope distinctions in sentences containing definition descriptions, in particular with respect to negation. The binary quantifier $I$ provides a means of formalising a theory of definite descriptions while building scope distinctions directly into the notation.

\section{The Binary Quantifier $I$ in \textbf{IPF}} 
Principle \rf{4} occupies common ground between positive and negative free logic, and so the introduction rule for the binary quantifier $I$ of \citep{kurbisiotaI} is good for the present system, too: 

\begin{prooftree}
\AxiomC{$F^x_t$}
\AxiomC{$G^x_t$}
\AxiomC{$\exists !t$}
\AxiomC{$\underbrace{[F_a^x]^i, \ [\exists !a]^j}$}
\noLine
\UnaryInfC{$\Pi$}
\noLine
\UnaryInfC{$a=t$}
\RightLabel{$_{i, j}$}
\LeftLabel{$II: \qquad$}
\QuaternaryInfC{$Ix[F, G]$}
\end{prooftree}

\noindent where $a$ is a fresh parameter, that is to say $a$ is different from $t$, does not occur in $F$ or $G$, nor in any undischarged assumption in $\Pi$ except $F_a^x$ and $\exists ! a$. 
\bigskip

\noindent The elimination rules for $I$ of \citep{kurbisiotaI} need to be adjusted. As $Ix[F, G]$ may be true even if no unique $F$ exists, they require the additional premise that the $F$ exists, which in the present symbolism is expressed by $Ix[F, \exists ! x]$. The following elimination rule for $I$ captures \rf{3}: 

\begin{prooftree}
\AxiomC{$Ix[F, G]$}
\AxiomC{$I x[F, \exists !x]$}
\AxiomC{$\underbrace{[F_a^x]^i, \ [G_a^x]^j, \ [\exists !a]^k}$}
\noLine
\UnaryInfC{$\Pi$}
\noLine
\UnaryInfC{$C$}
\LeftLabel{${I E^{1p}}': \quad$}
\RightLabel{$_{i, j, k}$}
\TrinaryInfC{$C$}
\end{prooftree}

\noindent where $a$ is a fresh parameter, that is to say $a$ does not occur in $F$, $G$ or $C$, nor in any undischarged assumptions of $\Pi$ except $F_a^x$, $G_a^x$ and $\exists ! a$, and it is not free in $F$ or $G$.\bigskip 

\noindent Although straightforward and convenient for practical purposes, from the proof-theoretic perspective this rule is unsatisfactory, as will be shown shortly. A modified version fares better. This explains why its label carries a prime.

One of the more idiosyncratic features of definite descriptions in positive free logic is that the uniqueness of the $F$ is also only consequent upon its existence, and so the second elimination rule for the binary quantifier of \citep{kurbisiotaI} would require the additional premise $Ix[F, \exists !x]$, too. This addition, however, would make the first premise redundant, and so in the present context we are left with: 

\begin{prooftree}
\AxiomC{$Ix[F, \exists !x]$}
\AxiomC{$\exists !t_1$}
\AxiomC{$\exists ! t_2$}
\AxiomC{$F_{t_1}^x$}
\AxiomC{$F_{t_2}^x$}
\LeftLabel{${I E^{2p}}': \quad$}
\QuinaryInfC{$t_1=t_2$}
\end{prooftree}

\noindent This rule, too, is going to be modified slightly, which explains the prime. 

The problem with ${IE^{1p}}'$ is that there is no general way of removing a formula of the form $Ix[F, \exists !x]$ from a deduction that is concluded by $II$ and the second premise of ${IE^{1p}}'$: 

\begin{prooftree}
\AxiomC{$I x[F, G]$}
\AxiomC{$F^x_t$}
\AxiomC{$\exists ! t$}
\AxiomC{$\exists !t$}
\AxiomC{$\underbrace{[F_z^x]^i, \ [\exists !z]^j}$}
\noLine
\UnaryInfC{$\Pi_1$}
\noLine
\UnaryInfC{$z=t$}
\RightLabel{$_{i, j}$}
\QuaternaryInfC{$I x[F, \exists ! x]$}
\AxiomC{$\underbrace{[F_z^x]^k, \ [G_z^x]^l, \ [\exists !z]^m}$}
\noLine
\UnaryInfC{$\Pi_2$}
\noLine
\UnaryInfC{$C$}
\RightLabel{$_{k, l, m}$}
\TrinaryInfC{$C$}
\end{prooftree}

\noindent This may not be a dramatic shortcoming: in intuitionist negative free logic, we cannot expect to be able to remove formulas of the form $\exists !t$ that have been concluded by rules of strictness and are used as the existence premises of $=I^n$, $\exists I$ or $\forall E$. We have a similar situation here, and so it would make sense not to consider ${IE^{1p}}'$ as an elimination rule for $I x[F, \exists !x]$, and not to count formulas that are concluded by $II$ and used as the second premise of ${IE^{1p}}'$ as maximal. 

Contrary to the situation in intuitionist negative free logic, however, it is possible to do better in $\mathbf{IPF}$. ${IE^{1p}}'$ can be reformulated in such a way that the offensive formulas $I x[F, \exists !x]$ may be removed from deductions. This comes at a cost in the complexity of the rules, so for the practical purpose of carrying out deductions it is often useful to keep the simpler rule in mind and apply it instead. The desired modification is achieved by replacing the premise $I x[F, \exists !x]$ of ${IE^{1p}}'$ by the conditions under which it may be derived as specified by $II$: 

\begin{prooftree}
\AxiomC{$Ix[F, G]$}
\AxiomC{$F^x_t$}
\AxiomC{$\exists ! t$}
\AxiomC{$\underbrace{[F_a^x]^{i_1}, \ [\exists ! a]^{i_2}}$}
\noLine
\UnaryInfC{$\Pi$}
\noLine
\UnaryInfC{$a=t$}
\AxiomC{$\underbrace{[F_b^x]^{i_3}, \ [G_b^x]^{i_4}, \ [\exists !b]^{i_5}}$}
\noLine
\UnaryInfC{$\Sigma$}
\noLine
\UnaryInfC{$C$}
\LeftLabel{$IE^{1p}: \quad$}
\RightLabel{$_{i_1\ldots i_5}$}
\QuinaryInfC{$C$}
\end{prooftree}

\noindent where $a$ and $b$ are fresh: $a$ is different from $t$, does not occur in $F$ or $G$, nor in any undischarged assumptions of $\Pi$ except $F_a^x$ and $\exists !a$; and $b$ does not occur in $F$, $G$, $C$, nor in any undischarged assumptions of $\Sigma$ except $F_b^x$, $G_b^x$ and $\exists ! b$.\bigskip

\noindent Replacing ${IE^{1p}}'$  by $IE^{1p}$ requires the addition of a further elimination rule for formulas of the form $I x[F, \exists !x]$: 

\begin{prooftree}
\AxiomC{$Ix[F, \exists !x]$}
\AxiomC{$\underbrace{[F_a^x]^i, [\exists !a]^j}$}
\noLine
\UnaryInfC{$\Pi$}
\noLine
\UnaryInfC{$C$}
\LeftLabel{$IE^{3p}: \quad$}
\RightLabel{$_{i, j}$}
\BinaryInfC{$C$}
\end{prooftree}

\noindent where $a$ is fresh: it does not occur in $F$ or $C$ nor in any undischarged assumptions of $\Pi$ except $F_a^x$ and $\exists ! a$.\bigskip

\noindent This is because $IE^{1p}$ makes stronger demands on the conditions under which it is applicable than ${IE^{1p}}'$: it requires the conditions under which the second premise of the latter may be derived, rather than just its assumption. ${IE^{1p}}'$ may therefore be applied under conditions where $IE^{1p}$ is not applicable, namely when $Ix[F, \exists !x]$ is assumed rather than deduced. $IE^{3p}$ restores the balance: it is effectively the special case of ${IE^{1p}}'$ where $G$ is $\exists!$.\footnote{It may be objected that in determining the elimination rules for $I$, I have not followed any of the methods for `reading off' elimination rules from introduction rules that may be found in the literature (see, e.g., \citep{prawitzmeaningandcompleteness}, \citep{schroederheisternatural}, \citep{nilsharmony}, \citep{nissimnote}, \cite{readGEharmony}, \cite{kurbisproofandfalsity}). I have instead looked for proof-theoretically satisfactory rules of inference by transposing axioms equivalent to a prominent theory of definite descriptions into rules of natural deduction with an eye on proving a normalisation theorem. A discussion whether this disqualifies the present approach in the eyes of some inferentialists is a broader question and goes beyond the scope of this paper. A decision must be be left to the reader.}

\begin{landscape} 
${IE^{1p}}'$ and the pair $IE^{1p}$ and $IE^{3p}$ are interderivable, given $II$ and ${IE^{2p}}'$: 

\noindent (a) $IE^{3p}$ is the special case of ${IE^{1p}}'$ where both premises are $I x[F, \exists ! x]$, but listed only once. Given $F_t^x$, $\exists ! t$ and a deduction $\Pi$ of $a=t$ from $F_a^x$ and $\exists ! a$, by $II$ derive $Ix[F, \exists !x]$, using $\exists ! t$ twice to make up the required number of premises, and so by ${IE^{1p}}'$, now using $Ix [F, \exists ! x]$ twice, derive $C$. 

\noindent (b) Given $Ix[F, G]$, $Ix[F, \exists ! x]$ and a deduction $\Pi$ of $C$ from $F_a^x$, $G_a^x$ and $\exists !a$, apply ${IE^{2p}}'$ to $Ix[F, \exists ! x]$ and assumptions $F_b^x$, $F_c^x$, $\exists ! b$ and $\exists ! c$, where $b, c$ are fresh and different, to derive $b=c$; using $F_c^x$ and $\exists ! c$ once more as premises, apply $IE^{1p}$ to derive $C$; apply ${IE^{3p}}$ to discharge assumptions $F_c^x$ and $\exists ! c$. See the construction below:  

\begin{prooftree}
\AxiomC{$Ix[F, \exists !x]$}
\AxiomC{$Ix[F, G]$}
\AxiomC{$[F_c^x]^k$}
\AxiomC{$[\exists ! c]^j$}
\AxiomC{$Ix[F, \exists ! x]$} 
\AxiomC{$[\exists ! b]^{i_1}$}
\AxiomC{$[\exists ! c]^j$}
\AxiomC{$[F_b^x]^{i_2}$}
\AxiomC{$[F_c^x]^k$}
\RightLabel{$_{{IE^{2p}}'}$}
\QuinaryInfC{$b=c$}
\AxiomC{$\underbrace{[F_a^x]^{i_3}, \ [G_a^x]^{i_4}, [\exists !a]^{i_5}}$}
\noLine
\UnaryInfC{$\Pi$}
\noLine
\UnaryInfC{$C$}
\RightLabel{$_{i_1\ldots i_5 \ IE^{1p}}$}
\QuinaryInfC{$C$}
\RightLabel{$_{j, k\ IE^{3p}}$}
\BinaryInfC{$C$}
\end{prooftree}

\noindent Thus $II$, $IE^{1p}$, ${IE^{2p}}'$ and $IE^{3p}$ capture \rf{3} and \rf{4}, just as well as do $II$ and ${IE^{1p}}'$. But we're not quite there yet. 

As shown by \rf{1}, identity sometimes carries some of the characteristics of existence, and we need additional rules to ensure this. They are more or less ${IE^{2p}}'$ and $IE^{3p}$ with $\exists!$ replaced by an identity: 

\begin{prooftree} 
\AxiomC{$Ix[F, x=t_2]$}
\AxiomC{$\exists !t_1$}
\AxiomC{$\exists ! t_2$}
\AxiomC{$F_{t_1}^x$}
\LeftLabel{${I E^{4p}}': \quad$}
\QuaternaryInfC{$t_1=t_2$}
\end{prooftree} 

\begin{prooftree} 
\AxiomC{$Ix[F, x=t]$}
\AxiomC{$\exists !t$}
\AxiomC{$\underbrace{[F_a^x]^i, [\exists !a]^j}$}
\noLine
\UnaryInfC{$\Pi$}
\noLine
\UnaryInfC{$C$}
\LeftLabel{$IE^{5p}: \quad$}
\RightLabel{$_{i, j}$}
\TrinaryInfC{$C$}
\end{prooftree}

\noindent Notice the missing premise $F_{t_2}^x$ in ${IE^{4p}}'$ and the additional premise $\exists !t$ in $IE^{5p}$. Like ${IE^{2p}}'$, ${IE^{4p}}'$ will be slightly modified, hence the prime. To this issue, we turn next. 
\end{landscape} 

It is possible to avoid occurrences of identities that are concluded by ${IE^{2p}}'$ and used as the major premise in Leibniz's Law by replacing the former rule with a slight reformulation that absorbs a step by the latter: 

\begin{prooftree}
\AxiomC{$Ix[F, \exists !x]$}
\AxiomC{$\exists !t_1 \qquad \exists ! t_2$}
\AxiomC{$F_{t_1}^x \qquad F_{t_2}^x$}
\AxiomC{$A_{t_1}^x$}
\LeftLabel{$IE^{2p}: \quad$}
\QuaternaryInfC{$A_{t_2}^x$}
\end{prooftree}

\noindent where $A$ is an atomic formula.\bigskip  

\noindent An induction over the complexity of formulas shows that $I E^{2p}$ is derivable for formulas $A$ of any degree. ${IE^{2p}}'$ and $I E^{2p}$ are interderivable in virtue of the rules for identity. To derive ${IE^{2p}}'$ from $I E^{2p}$, let $A_{t_1}^x$ be $t_1=t_1$; for the converse, apply Leibniz's Law to the conclusion of ${IE^{2p}}'$. 

Finally, we do the same with ${IE^{4p}}'$: 

\begin{prooftree} 
\AxiomC{$Ix[F, x=t_2]$}
\AxiomC{$\exists !t_1$}
\AxiomC{$\exists ! t_2$}
\AxiomC{$F_{t_1}^x$}
\AxiomC{$A_{t_1}^x$}
\LeftLabel{$IE^{4p}: \quad$}
\QuinaryInfC{$A_{t_2}^x$}
\end{prooftree} 

\noindent where $A$ is an atomic formula.\bigskip  

\noindent An argument similar to the one given in the previous case shows that $IE^{4p}$ and ${IE^{4p}}'$ are interderivable and an induction that the rules is derivable for formulas $A$ of any degree. 

For practical purposes ${IE^{2p}}'$ and ${IE^{4p}}'$ are more convenient than $IE^{2p}$ and $IE^{4p}$, but for proof-theoretic purposes the latter are more interesting. So let $\mathbf{IPF}^I$ be $\mathbf{IPF}$ extended by the binary quantifier $I$ governed by the rules $II$, $IE^{1p}, IE^{2p}, IE^{3p}, IE^{4p}, IE^{5p}$. In the next section we will prove a normalisation theorem for this system. 
 
To close this section, it may not come amiss to illustrate the use of $I$ with a few examples. For simplicity, throughout I'll use ${IE^{2p}}'$ and ${IE^{4p}}'$ instead of $IE^{2p}$ and $IE^{4p}$. Applications of the unfamiliar rules for $I$ will be marked explicitly, for easier readability. The deductions also show that $\mathbf{IPF}^I$ provides an adequate reconstruction of a minimal theory of definite descriptions within positive free logic. The examples correspond to characteristic theses of a minimal theory of definite descriptions formalised with a term-forming operator and axiomatised by \rf{LL}. The correspondence between $\mathbf{IPF}^I$ and $\mathbf{IPF}^\iota$ is of course not perfect, as the latter system does not have scope distinctions. A closer comparison of the two systems is the subject of section 6. 

\lbp{10}{$\ast 10$}{$\exists x (F\land\forall y(F_y^x\rightarrow y=x)\land G)\vdash Ix[F, G]$}

\noindent See \citep[91]{kurbisiotaI}: the deduction given there only appeals to the introduction rule for the binary quantifier, so remains correct in $\mathbf{IPF}^I$. 

\begin{landscape} 
\lbp{11}{$\ast 11$}{$Ix[F, G], Ix[F, \exists !x] \vdash  \exists x (F\land\forall y(F_y^x\rightarrow y=x)\land G)$}

\begin{prooftree}
\def\defaultHypSeparation{\hskip .01in}
\AxiomC{$Ix[F, \exists !x]$}
\AxiomC{$Ix[F, G]$}
\AxiomC{$[F_d^x]^8$}
\AxiomC{$[\exists ! d]^9$}
\AxiomC{$Ix[F, \exists !x]$}
\AxiomC{$[\exists !c]^3$}
\AxiomC{$[\exists ! d]^9$}
\AxiomC{$[F_c^x]^4$}
\AxiomC{$[F_d^x]^8$}
\RightLabel{$_{{IE^{2p}}'}$}
\QuinaryInfC{$c=d$}
\AxiomC{$[F_b^x]^5$}
\AxiomC{$Ix[F, \exists !x]$}
\AxiomC{$[\exists !a]^2$}
\AxiomC{$[\exists ! b]^6$}
\AxiomC{$[F_a^x]^1$}
\AxiomC{$[F_b^x]^5$}
\RightLabel{$_{{IE^{2p}}'}$}
\QuinaryInfC{$a=b$}
\RightLabel{$_1$}
\UnaryInfC{$F_a^x\rightarrow a=b$}
\RightLabel{$_2$}
\UnaryInfC{$\forall y(F_y^x\rightarrow y=b)$}
\AxiomC{$[G_b^x]^7$}
\BinaryInfC{$\forall y(F_y^x\rightarrow y=b)\land G_b^x$}
\BinaryInfC{$F_b^x\land \forall y(F_y^x\rightarrow y=b)\land G_b^x$}
\UnaryInfC{$\exists x(F\land \forall y(F_y^x\rightarrow y=b)\land G)$}
\RightLabel{$_{3, 4, 5, 6, 7 \ IE^{1p}}$}
\QuinaryInfC{$\exists x(F\land \forall y(F_y^x\rightarrow y=b)\land G)$}
\RightLabel{$_{8, 9 \ IE^{3p}}$}
\BinaryInfC{$\exists x(F\land \forall y(F_y^x\rightarrow y=b)\land G)$}
\end{prooftree}

\lbp{12}{$\ast 12$}{$Ix[F, \exists !x]\vdash \exists y\forall x(F\leftrightarrow x=y)$}

\begin{prooftree}
\AxiomC{$Ix[F, \exists ! x]$}
\AxiomC{$Ix[F, \exists ! x]$}
\AxiomC{$[\exists !a]^3$}
\AxiomC{$[\exists !b]^4$}
\AxiomC{$[F_a^x]^1$}
\AxiomC{$[F_b^x]^5$}
\RightLabel{$_{{IE^{2p}}'}$}
\QuinaryInfC{$a=b$}
\AxiomC{$[F_b^x]^5$}
\AxiomC{$[a=b]^2$}
\BinaryInfC{$F_a^x$}
\RightLabel{$_{1, 2}$}
\BinaryInfC{$F_a^x\leftrightarrow a=b$}
\RightLabel{$_3$}
\UnaryInfC{$\forall x(F\leftrightarrow x=b)$}
\AxiomC{$[\exists !b]^4$}
\BinaryInfC{$\exists y\forall x(F\leftrightarrow x=y)$}
\RightLabel{$_{4, 5 \ IE^{3p}}$}
\BinaryInfC{$\exists y\forall x(F\leftrightarrow x=y)$}
\end{prooftree}

\newpage 

\lbp{13}{$\ast 13$}{$\exists y\forall x(F\leftrightarrow x=y)\vdash Ix[F, \exists !x]$}

\begin{prooftree}
\AxiomC{$\exists y\forall x(F\leftrightarrow x=y)$}
\AxiomC{$[\forall x(F\leftrightarrow x=a)]^3$}
\AxiomC{$[\exists ! a]^4$}
\BinaryInfC{$F_a^x\leftrightarrow a=a$}
\AxiomC{$a=a$}
\BinaryInfC{$F_a^x$}
\AxiomC{$[\exists ! a]^4$}
\AxiomC{$[\exists ! a]^4$}
\AxiomC{$[\forall x(F\leftrightarrow x=a)]^3$}
\AxiomC{$[\exists ! b]^1$}
\BinaryInfC{$F_b^x\leftrightarrow b=a$}
\AxiomC{$[F_b^x]^2$}
\BinaryInfC{$b=a$}
\RightLabel{$_{1, 2 \ II}$}
\QuaternaryInfC{$Ix[F, \exists !x]$}
\RightLabel{$_{3, 4}$}
\BinaryInfC{$Ix[F, \exists !x]$}
\end{prooftree}

\lbp{14}{$\ast 14$}{$\vdash Ix[F, \exists !x]\leftrightarrow \exists y\forall x(F\leftrightarrow x=y)$}

\noindent From \rf{12} and \rf{13} by $\leftrightarrow I$. 

\lbp{15}{$\ast 15$}{$Ix[F, \exists !x]\vdash \exists yIx[F, x=y]$}

\begin{prooftree} 
\AxiomC{$Ix[F, \exists !x]$}
\AxiomC{$[\exists !a]^4$} 
\AxiomC{$[F_a^x]^3$}
\AxiomC{$a=a$} 
\AxiomC{$[\exists !a]^4$}
\AxiomC{$Ix[F, \exists !x]$} 
\AxiomC{$[\exists !b]^1$} 
\AxiomC{$[\exists !a]^4$}
\AxiomC{$[F_b^x]^2$}
\AxiomC{$[F_a^x]^3$}
\RightLabel{$_{{IE^{2p}}'}$}
\QuinaryInfC{$b=a$} 
\RightLabel{$_{1, 2 \ II}$}
\QuaternaryInfC{$Ix[F, x=a$]}
\BinaryInfC{$\exists yIx[F, x=y$]}
\RightLabel{$_{3, 4 \ IE^{3p}}$}
\BinaryInfC{$\exists yIx[F, x=y$]}
\end{prooftree} 

\lbp{16}{$\ast 16$}{$\exists yIx[F, x=y]\vdash Ix[F, \exists !x]$} 

\begin{prooftree} 
\AxiomC{$\exists yIx[F, x=y]$}
\AxiomC{$[Ix[F, x=a]]^5$} 
\AxiomC{$[\exists ! a]^6$}
\AxiomC{$[F_a^x]^3$} 
\AxiomC{$[\exists !a]^4$} 
\AxiomC{$[\exists !a]^4$} 
\AxiomC{$[Ix[F, x=a]]^5$} 
\AxiomC{$[\exists !b]^2$} 
\AxiomC{$[\exists !a]^4$} 
\AxiomC{$[F_b^x]^1$}
\RightLabel{$_{{IE^{4p}}'}$}
\QuaternaryInfC{$b=a$} 
\RightLabel{$_{1, 2 \ II}$}
\QuaternaryInfC{$Ix[F, \exists !x]$}
\RightLabel{$_{3, 4 \ IE^{5p}}$}
\TrinaryInfC{$Ix[F, \exists !x]$} 
\RightLabel{$_{5, 6}$} 
\BinaryInfC{$Ix[F, \exists !x]$} 
\end{prooftree} 

\lbp{17}{$\ast 17$}{$\vdash \exists yIx[F, x=y]\leftrightarrow Ix[F, \exists !x]$}

\noindent From \rf{15} and \rf{16} by $\leftrightarrow I$.\bigskip

\noindent \rf{17} corresponds to the special case of \rf{1} in a system with a term-forming $\iota$ operator where $t$ is a definite description. 

\lbp{18}{$\ast 18$}{$Ix[F, x=a], \exists !a\vdash Ix[F, \exists !x]$}

\noindent By omitting the final application of $\exists E$ in the proof of \rf{16}.\bigskip

\noindent \rf{18} corresponds to the application of Leibniz's Law in the derivation of \rf{5}. Thus a formula corresponding to \rf{LL} is derivable in $\mathbf{IPF}^I$ by adapting the proof of \rf{5}.  

The following two propositions correspond to $\forall E$ and $\exists I$ when $t$ is a definite description: 

\lbp{19}{$\ast 19$}{$\forall x G, Ix[F, \exists !x]\vdash Ix[F, G]$} 

\noindent See \citep[315]{kurbisiotaII}: the proof is exactly as the one given there. 

\lbp{20}{$\ast 20$}{$Ix[F, G], Ix[F, \exists !x]\vdash \exists xG$} 

\begin{prooftree}
\AxiomC{$Ix[F, \exists !x]$}
\AxiomC{$Ix[F, G]$}
\AxiomC{$[F^x_b]^5$}
\AxiomC{$[\exists ! b]^6$}
\AxiomC{$Ix[F, \exists !x]$}
\AxiomC{$[\exists !a]^1$}
\AxiomC{$[\exists !b]^6$}
\AxiomC{$[F_a^x]^2$}
\AxiomC{$[F_b^x]^5$}
\RightLabel{$_{{IE^{2p}}'}$}
\QuinaryInfC{$a=b$}
\AxiomC{$[G_c^x]^3$}
\AxiomC{$[\exists ! c]^4$}
\BinaryInfC{$\exists xG$}
\RightLabel{$_{1, 2, 3, 4 \ IE^{1p}}$}
\QuinaryInfC{$\exists xG$}
\RightLabel{$_{5, 6 \ IE^{3p}}$}
\BinaryInfC{$\exists xG$}
\end{prooftree}
\end{landscape}

\section{Normalisation for $\mathbf{IPF}^I$}
The \emph{major premise} of an elimination rule is the premise that contains the connective it governs in the general statement of the rule: they way elimination rules are written here, it is always their leftmost premise.\footnote{The terminology in this section follows \citep{troelstraschwichtenberg}.}

\begin{definition}[Maximal Formula]
\normalfont A \emph{maximal formula} is an occurrence of a formula in a deduction that is the conclusion of an introduction rule and major premise of an elimination rule. 
\end{definition} 

\noindent Call the rules $\lor E$, $\exists E$, $IE^{1p}$, $IE^{3p}$ and $IE^{5p}$ \emph{del-rules}. 

\begin{definition}[Segment, Length and Degree of a Segment, Maximal Segment]
\normalfont (a) A \emph{segment} is a sequence of two or more formula occurrences $C_1\ldots C_n$ in a deduction such that $C_1$ is not the conclusion of a del-rule, $C_n$ is not the minor premise of a del-rule, and for every $i<n$, $C_i$ is minor premise of a del-rule and $C_{i+1}$ its conclusion. 

\noindent (b) The \emph{length} of a segment is the number of formula occurrences of which it consists, its \emph{degree} is their degree. 

\noindent (c) A segment is \emph{maximal} if and only if its last formula is the major premise of an elimination rule.
\end{definition}

\begin{definition}[Normal Form]
\normalfont A deduction is in \emph{normal form} if and only if it contains neither maximal formulas nor maximal segments. 
\end{definition} 

\noindent \emph{Detour conversions} are methods for removing maximal formulas from deductions. \emph{Permutation conversions} decrease the length of maximal segments by permuting the application of the elimination rule to its last formula upwards. If its first formula was derived by an introduction rule, the procedure turns the maximal segment into a maximal formula. I refer to both kinds of conversions collectively as \emph{reduction steps}. 

\begin{definition}[Rank of a Deduction]
\normalfont The \emph{rank} of a deduction $\Pi$ is the pair $\langle d, l\rangle$ where $d$ is the highest degree of any maximal formula or segment in $\Pi$, and $l$ is the sum of the number of maximal formulas plus the sum of the lengths of all maximal segments in $\Pi$. If there are no maximal formulas or segments in $\Pi$, let its rank be $\langle 0, 0\rangle$.
\end{definition} 

\noindent Ranks are ordered lexicographically: $\langle d, l\rangle< \langle d', l'\rangle$ iff either $d<d'$ or $d=d'$ and $l<l'$. 

Reduction steps for the connectives of $\mathbf{IPF}$ are straightforward by adapting those given by Prawitz \citep{prawitznaturaldeduction}. We state without giving the details: 

\begin{theorem}
Any deduction in $\mathbf{IPF}$ can be brought into normal form. 
\end{theorem} 

\noindent \emph{Proof} by induction over the rank of deductions in \textbf{IPF}: applying a reduction step to a suitably chosen maximal formula or maximal segment of highest degree reduces the rank of a deduction. 

\begin{landscape}
\noindent To prove that deductions in $\mathbf{IPF}^I$ normalise, we need to add detour conversions to remove maximals formulas of the form $Ix[F, G]$, where in case the maximal formula is eliminated by $IE^{1p}$, $G$ can be any formula, while if it is eliminated by $IE^{2p}$ or $IE^{3p}$ it is $\exists !x$, and if it is eliminated by $IE^{4p}$ or $IE^{5p}$ it is $x=t$. I will only give detour conversions, the permutation conversions being standard.\bigskip 

\noindent \emph{Detour Conversions for the Binary Quantifier I} 

\noindent In each case, the conditions on parameters and the fact that every deduction can be transformed into one in which each parameter is the parameter of exactly one application of a rule for a quantifier ensure that the deduction remains correct after the conversion.\bigskip

\noindent 1. The maximal formula is major premise of $IE^{1p}$. Replace the deduction to the left of $\leadsto$ with the one on its right:

\begin{center} 
\AxiomC{$\Pi_1$}
\noLine
\UnaryInfC{$F^x_{t_1}$}
\AxiomC{$\Pi_2$}
\noLine
\UnaryInfC{$G^x_{t_1}$}
\AxiomC{$\Pi_3$}
\noLine
\UnaryInfC{$\exists !t_1$}
\AxiomC{$\underbrace{[F_a^x]^i, \ [\exists !a]^j}$}
\noLine
\UnaryInfC{$\Pi_4$}
\noLine
\UnaryInfC{$a=t_1$}
\RightLabel{$_{i, j}$}
\QuaternaryInfC{$Ix[F, G]$}
\AxiomC{$\Sigma_1$}
\noLine
\UnaryInfC{$F^x_{t_2}$}
\AxiomC{$\Sigma_2$}
\noLine
\UnaryInfC{$\exists ! t_2$}
\AxiomC{$\underbrace{[F_b^x]^{k_1}, \ [\exists ! b]^{k_2}}$}
\noLine
\UnaryInfC{$\Sigma_3$}
\noLine
\UnaryInfC{$a=t_2$}
\AxiomC{$\underbrace{[F_c^x]^{k_3}, \ [G_c^x]^{k_4}, \ [\exists !c]^{k_5}}$}
\noLine
\UnaryInfC{$\Xi$}
\noLine
\UnaryInfC{$C$}
\RightLabel{$_{k_1\ldots k_5}$}
\QuinaryInfC{$C$}
\DisplayProof\qquad$\leadsto$\qquad
\AxiomC{$\underbrace{\stackon{[F_{t_1}^x]}{\Pi_1}, \ \stackon{[G_{t_1}^x]}{\Pi_2}, \ \stackon{[\exists !t_1]}{\Pi_3}}$}
\noLine
\UnaryInfC{$\Xi_{t_1}^c$}
\noLine
\UnaryInfC{$C$}
\DisplayProof
\end{center}

\noindent 2. The maximal formula is major premise of $IE^{2p}$. Replace the deduction to the left of $\leadsto$ with the one on its right, where the double line indicates the steps needed to derive the symmetry of identity: 

\begin{center} 
\AxiomC{$\Pi_1$}
\noLine
\UnaryInfC{$F^x_{t_1}$}
\AxiomC{$\Pi_2$}
\noLine
\UnaryInfC{$G^x_{t_1}$}
\AxiomC{$\Pi_3$}
\noLine
\UnaryInfC{$\exists !t_1$}
\AxiomC{$\underbrace{[F_a^x]^i, \ [\exists !a]^j}$}
\noLine
\UnaryInfC{$\Xi$}
\noLine
\UnaryInfC{$a=t_1$}
\RightLabel{$_{i, j}$}
\QuaternaryInfC{$Ix[F, \exists !x]$}
\AxiomC{$\stackon{\exists !t_2}{\Sigma_1} \qquad \stackon{\exists ! t_3}{\Sigma_2}$}
\AxiomC{$\stackon{F_{t_2}^x}{\Sigma_3} \qquad \stackon{F_{t_3}^x}{\Sigma_4}$}
\AxiomC{$\Sigma_5$}
\noLine
\UnaryInfC{$A_{t_2}^x$}
\QuaternaryInfC{$A_{t_3}^x$}
\DisplayProof\qquad$\leadsto$\qquad
\AxiomC{$\underbrace{\stackon{[F_{t_3}^x]}{\Sigma_4}, \ \stackon{[\exists !t_3]}{\Sigma_2}}$}
\noLine
\UnaryInfC{$\Xi_{t_3}^a$}
\noLine
\UnaryInfC{$t_3=t_1$}
\doubleLine
\UnaryInfC{$t_1=t_3$}
\AxiomC{$\underbrace{\stackon{[F_{t_2}^x]}{\Sigma_3}, \ \stackon{[\exists !t_2]}{\Sigma_1}}$}
\noLine
\UnaryInfC{$\Xi_{t_2}^a$}
\noLine
\UnaryInfC{$t_2=t_1$}
\BinaryInfC{$t_2=t_3$} 
\AxiomC{$\Sigma_5$}
\noLine
\UnaryInfC{$A_{t_2}^x$}
\BinaryInfC{$A_{t_3}^x$}
\DisplayProof
\end{center}

\noindent This last reduction step is quite interesting. It is unusual for reduction steps to require applications of rules of inference for a different connective, in this case the rules for identity. As identity occurs in $II$, this is what we should expect. If normalisation was possible without applying the rules for identity, it would appear that the rules governing $I$ are in harmony independently of identity, so that there should be rules governing $I$ that do not appeal to it. But this is impossible in the present framework, as we cannot express uniqueness without using identity.\bigskip 

\noindent 3. The maximal formula is major premise of $IE^{3p}$. Replace the deduction to the left of $\leadsto$ with the one on its right: 

\begin{center} 
\AxiomC{$\Pi_1$} 
\noLine
\UnaryInfC{$F^x_t$}
\AxiomC{$\Pi_2$}
\noLine
\UnaryInfC{$\exists ! t$}
\AxiomC{$\Pi_3$}
\noLine
\UnaryInfC{$\exists !t$}
\AxiomC{$\underbrace{[F_a^x]^i, \ [\exists !a]^j}$}
\noLine
\UnaryInfC{$\Pi_4$}
\noLine
\UnaryInfC{$a=t$}
\RightLabel{$_{i, j}$}
\QuaternaryInfC{$Ix[F, \exists ! x]$}
\AxiomC{$\underbrace{[F_b^x]^k, \ [\exists !b]^l}$}
\noLine
\UnaryInfC{$\Xi$}
\noLine
\UnaryInfC{$C$}
\RightLabel{$_{k, l}$}
\BinaryInfC{$C$}
\DisplayProof\qquad$\leadsto$\qquad
\AxiomC{$\underbrace{\stackon{[F_t^x]}{\Pi_1}, \ \stackon{[\exists !t]}{\Pi_2}}$}
\noLine
\UnaryInfC{$\Xi_t^b$}
\noLine
\UnaryInfC{$C$}
\DisplayProof
\end{center}

\noindent Alternatively, we could replace $t$ by $b$ in $\Pi_1$ and $\Pi_2$ or $\Pi_3$, thereby concluding $F_b^x$ and $\exists !b$, using them to conclude the respective open assumptions of $\Xi$, and continue on to conclude $C$.\bigskip

\noindent 4. The maximal formula is major premise of $IE^{4p}$. Replace the deduction to the left of $\leadsto$ with the one on its right, where the double line indicates the steps needed to derive the symmetry of identity: 

\begin{center}
\AxiomC{$\Pi_1$}
\noLine
\UnaryInfC{$F^x_{t_1}$}
\AxiomC{$\Pi_2$}
\noLine
\UnaryInfC{$t_1=t_2$}
\AxiomC{$\Pi_3$}
\noLine
\UnaryInfC{$\exists !t_1$}
\AxiomC{$\underbrace{[F_a^x]^i, \ [\exists !a]^j}$}
\noLine
\UnaryInfC{$\Xi$}
\noLine
\UnaryInfC{$a=t_1$}
\RightLabel{$_{i, j}$}
\QuaternaryInfC{$Ix[F, x=t_2]$}
\AxiomC{$\Sigma_1$}
\noLine
\UnaryInfC{$\exists !t_3$}
\AxiomC{$\Sigma_2$}
\noLine
\UnaryInfC{$\exists !t_2$}
\AxiomC{$\Sigma_3$}
\noLine
\UnaryInfC{$F_{t_3}^x$}
\AxiomC{$\Sigma_4$}
\noLine
\UnaryInfC{$A_{t_2}^x$}
\QuinaryInfC{$A_{t_3}^x$}
\DisplayProof\qquad$\leadsto$\qquad
\AxiomC{$\Pi_2$} 
\noLine
\UnaryInfC{$t_1=t_2$}
\AxiomC{$\underbrace{\stackon{[F_{t_3}^x]}{\Sigma_3}, \ \stackon{[\exists !t_3]}{\Sigma_1}}$}
\noLine
\UnaryInfC{$\Xi_{t_3}^a$}
\noLine
\UnaryInfC{$t_3=t_1$}
\BinaryInfC{$t_3=t_2$} 
\doubleLine
\UnaryInfC{$t_2=t_3$}
\AxiomC{$\Sigma_4$}
\noLine
\UnaryInfC{$A_{t_2}^x$}
\BinaryInfC{$A_{t_3}^x$}
\DisplayProof
\end{center} 

\noindent In this reduction step, too, appeal is made to the rules for identity: as before, this is not surprising, this time because an identity is a subformula of the maximal formula to be removed.\bigskip

\noindent 5. The maximal formula is major premise of $IE^{5p}$. Replace the deduction to the left of $\leadsto$ with the one on its right, where the double line indicates the steps needed to derive the symmetry of identity: 

\begin{center} 
\AxiomC{$\Pi_1$} 
\noLine
\UnaryInfC{$F^x_{t_1}$}
\AxiomC{$\Pi_2$}
\noLine
\UnaryInfC{$t_1=t_2$}
\AxiomC{$\Pi_3$}
\noLine
\UnaryInfC{$\exists !t_1$}
\AxiomC{$\underbrace{[F_a^x]^i, \ [\exists !a]^j}$}
\noLine
\UnaryInfC{$\Pi_4$}
\noLine
\UnaryInfC{$a=t_1$}
\RightLabel{$_{i, j}$}
\QuaternaryInfC{$Ix[F, x=t_2]$}
\AxiomC{$\Sigma$}
\noLine
\UnaryInfC{$\exists t_2$}
\AxiomC{$\underbrace{[F_b^x]^k, \ [\exists !b]^l}$}
\noLine
\UnaryInfC{$\Xi$}
\noLine
\UnaryInfC{$C$}
\RightLabel{$_{k, l}$}
\TrinaryInfC{$C$}
\DisplayProof\qquad$\leadsto$\qquad
\AxiomC{$\underbrace{\stackon{F_{t_1}^x}{\Pi_1}, \ \stackon{\exists !t_1}{\Pi_3}}$}
\noLine
\UnaryInfC{$\Xi_{t_1}^x$}
\noLine
\UnaryInfC{$C$}
\DisplayProof
\end{center}

\noindent This completes the detour conversions for maximal formulas of the form $Ix[F, G]$, and we are ready to prove: 

\begin{theorem}
Any deduction in $\mathbf{IPF}^I$ can be brought into normal form. 
\end{theorem}

\noindent \emph{Proof} by induction over the rank for deductions. Applying a reduction step to a maximal formula or maximal segment of highest degree such that no such segments stand above it reduces the rank of the deduction. 
\end{landscape}

\section{The Binary Quantifier $I$ and the Term-Forming $\iota$ Operator in $\mathbf{IPF}$} 
For reasons sufficiently indicated in the previous sections, as $\mathbf{IPF}^\iota$ lacks a means for drawing scope distinctions, I will not compare the full systems $\mathbf{IPF}^I$ and $\mathbf{IPF}^\iota$. There is, for instance, no direct, straightforward translation from one to the other. Instead, I impose two restrictions on both systems and compare the results: (1) the $G$ in $Ix[F, G]$ is either $\exists!$ or $=$, and correspondingly, $\iota$-terms occur only after $\exists!$ or to the left or right of identity (and not both); (2) Leibniz's Law is restricted to constants and parameters. (1) excludes formulas such as $\neg G(\iota xF)$, where questions of scope arise, and absolves us from considering nested binary quantifiers. It also permits us to regard $Ix[A, \exists !x]$ and $\exists !\iota xA$ and $Ix[A, x=t]$ and $\iota xA=t$ as notational variants, as was done in a previous paper \citep[Sec 4]{kurbisiotaII}. (2) is justified with an eye to extending $\mathbf{IPF}^I$ by modal operators, in which case restrictions of Leibniz' Law are mandatory if non-rigid terms are considered.\footnote{An even stricter approach is followed by Fitting and Mendelsohn, who only allow variables to occur to the left and right of identity \citep[Ch 7]{mendelsohnfitting}; in fact, the atomic formulas of their system are formed only from predicate letters and variables \citep[81]{mendelsohnfitting}. To form a formula with a name, function symbol or definite description requires a predicate formed by predicate abstraction \citep[196f, 248f]{mendelsohnfitting}. Variables are interpreted rigidly, all other terms may be rigid or non-rigid. Leibniz' Law holds unrestrictedly only for variables.} 

There is one respect in which (1) does not present much of a restriction of $\mathbf{IPF}^\iota$ at all. In $\mathbf{IPF}^\iota$ the logical force of formulas of the form $G(\iota xF)$ is determined entirely by that of formulas of the form $Gt$ and $t=\iota xF$: $G(\iota xF)$ cannot be used in a deduction unless there is also a formula $t=\iota xF$, for some term $t$. The consequences of $G(\iota xF)$ and the conditions under which it may be inferred are then specified by \rf{LL} and Leibniz's Law. Thus instead of considering  $G(\iota xF)$, we may consider $G(t)\land t=\iota xF$, for some term $t$, instead. In other respects (1) and (2) of course present significant restrictions. 


Let $\mathbf{IPF}^{IR}$ and $\mathbf{IPF}^{\iota R}$ be $\mathbf{IPF}^I$ and $\mathbf{IPF}^\iota$ with the respective versions of restrictions (1) and (2) imposed. We will show them to be equivalent. 

\rf{5} derives \rf{LL} from instances of \rf{3} and \rf{4} respecting restriction (1). \rf{10} and \rf{11} derive the notational variants of \rf{3} and \rf{4} in $\mathbf{IPF}^I$, and hence the required instances hold in  $\mathbf{IPF}^{IR}$. \rf{18} is the instance of Leibniz's Law corresponding to the one application of it to an $\iota$ term in \rf{5}. Any other application of Leibniz's Law in these deductions respects restriction (2). Thus changing notation and replacing formulas of the form $\iota xF=t$ in \rf{5} to their notational variant $Ix[F, x=t]$, it follows that \rf{LL} is derivable in $\mathbf{IPF}^{IR}$. Thus $\mathbf{IPF}^{\iota R}$ is a subsystem of $\mathbf{IPF}^{IR}$. It is worth checking that all six rules for the binary quantifier are used in the deductions. 

To show the converse, it suffices to observe that the notational variants of $II$, ${I E^{1p}}'$, ${\iota E^{2p}}'$, ${IE^{4p}}'$ and $IE^{5p}$ in $\mathbf{IPF}^{\iota R}$ are straightforward consequences of \rf{LL}. $\mathbf{IPF}^{IR}$ is thus a subsystem of $\mathbf{IPF}^{\iota R}$, and so we have:

\begin{theorem}
$\mathbf{IPF}^{IR}$ and $\mathbf{IPF}^{\iota R}$ are equivalent.
\end{theorem}

\section{Conclusion} 
The binary quantifier $I$ allows for the formalisation of sentences containing definite descriptions while respecting intuitive distinctions of scope without the need for introducing an additional syntactic means for representing them in the formal system. Its rules have desirable proof-theoretic properties: deductions in $\mathbf{IPF}^I$ normalise. A subsystem of $\mathbf{IPF}^I$ was shown to be equivalent to a system formalising the minimal theory of a term forming $\iota$ operator for definite descriptions within intuitionist positive free logic. This represents the more common approach to the formalisation of definite descriptions. 

Comparing $\mathbf{IPF}^I$ to a system that extends $\mathbf{IPF}^\iota$ by a device such as a $\lambda$ operator for predicate abstraction to mark scope must be left to another occasion. Classical systems of this kind found in the literature on free logic are not suitable to the present concerns. The system of \citep{lambertfreedef} is designed for negative free logic, while predicate abstraction formalised in \cite{lambertbencivengacomplexpred} carries existential import. Thus an expression of the form $\lambda xG(\iota xF)$ of the latter system extended by the $\iota$ operator does not correspond to $Ix[F, G]$, which does not carry existential import. 

More suitable systems may be found in the context of modal logic, such as those of \citep[Ch 9ff]{mendelsohnfitting} and \citep[Ch 19]{garsonmodallogic}. Indrzejczak has provided cut free sequent calculi for the former \citep{andrzejmodaldescription} as well as the latter \citep{andrzejexistencedefinedness}. These results inspire confidence that similarly proof-theoretically satisfactory systems may be provided for modal logics extended by the binary quantifier $I$. An extension of $\mathbf{IPF}^I$ by modal operators and a comparison of the result with these systems will be the focus of future investigations.\bigskip

\noindent \textbf{Acknowledgements.} I would like to thank Andrzej Indrzejczak for discussions of the proof-theory of definite descriptions and encouragement to write this paper. Two referees for this journal also made helpful comments on a previous version. Its final version was prepared while I was an Alexander von Humboldt fellow at the University of Bochum. 

\bigskip

\setlength{\bibsep}{0pt}
\bibliographystyle{chicago}
\bibliography{iota}

\end{document}